# Facility class Rayleigh beacon AO system for the 4.2m William Herschel Telescope[*]

René G.M. Rutten[a,b], Paul Clark[c], Richard M. Myers[c], Richard W. Wilson[c], Richard G. Bingham[d], Eric Emsellem[e], Thomas Gregory[a], Ronald A. Humphreys[c], Johan H. Knapen[a,f], Gil Moretto[c], Simon L. Morris[c], Gordon Talbot[a]

[a] Isaac Newton Group of Telescopes, Spain; [b] Netherlands Organization for Scientific Research, NL; [c] University of Durham, UK; [d] University College London, UK; [e] Observatoire de Lyon, France; [f] University of Hertfordshire, UK

## ABSTRACT

A rationale is presented for the use of a relatively low-altitude (15km) Rayleigh Laser Guide Star to provide partial adaptive optics correction across a large fraction of the sky on the 4.2m William Herschel Telescope. The scientific motivation in relation to the available instrumentation suite is discussed and supported by model performance calculations, based on observed atmospheric turbulence distributions at the site. The proposed implementation takes the form of a laser system, beam diagnostics, tip-tilt mirror and beacon launch telescope, together with a range-gated wavefront sensor and processing system. It is designed to operate in conjunction with the telescope's existing facility-class natural guide star AO system, NAOMI. Aspects of the proposed implementation are described as well as the technical features related to the system model and the error budget. In a separate paper the NAOMI AO system itself is presented. Other papers describe a demonstrator system and preliminary Rayleigh beacon wavefront sensing measurements at the site.

## INTRODUCTION

It is now well established that adaptive optics (AO) techniques allow ground-based observers to obtain high spatial resolution by correcting the image blurring introduced by the Earth's atmosphere. The resulting image sharpness with AO readily reaches the diffraction limit at near IR wavelengths, provided atmospheric turbulence conditions are good and a relatively bright point source for wavefront measurement is available. Higher spatial resolution in general provides important scientific advantages as has been best demonstrated by the *Hubble Space Telescope*. Achieving high spatial resolution not only carries the advantage of distinguishing finer structure and avoiding source confusion in dense fields, but also allows observations to reach significantly fainter as the sky background component reduces. For these reasons, AO instrumentation is being planned for nearly all large telescopes, and is at the heart of the future generation of extremely large telescopes. At the 4.2-m William Herschel Telescope (WHT) on the island La Palma, AO recently came to fruition with the commissioning of the common-user AO system, NAOMI, and an aggressive instrument development programme.

At present, the main practical limitation for AO is the availability of bright guide stars, point sources near the object of interest that are used to measure the wavefront distortions caused by atmospheric turbulence. For normal AO the percentage of sky that can be observed is limited to ~1% at visible wavelengths, which obviously severely restricts the

---

applicability of AO as a general-purpose tool. It is this limitation that has caused AO to produce fewer science results than one might have expected from its potential.

This limitation can largely be lifted by generating an artificial guide star using a laser launched alongside the telescope. Sky coverage for AO exploitation can thus be increased very significantly, approaching 100%. This opens up AO to virtually *all* areas of observational astronomy and to virtually *all* positions in the sky. In particular, it opens up the possibility of observing faint and extended sources, while enabling observations of large samples, unbiased by the fortuitous presence of nearby bright stars. For this reason a small number of telescopes currently have operational laser systems and several telescopes have plans to implement laser operation with their AO system[1].

The scientific exploitation of 4-m class telescopes is changing as a result of the availability of a new generation of 10-m class telescopes; so also for the WHT. The current instrumentation development strategy for the WHT focuses on two strands: (i) exploiting the wide field in the prime focus, and (ii) exploiting AO. The laser guide star (LGS) plans described in this paper fall into the second category. In preparation for the construction of a fully functioning laser guide star system laser launch experiments have been carried out at the WHT. These experiments are described by Clark et al. elsewhere in these proceedings.

## SCIENTIFIC RATIONALE FOR A RAYLEIGH LASER BEACON

The rationale for focussing on adaptive optics at the WHT may be summarised as follows. Firstly, most types of astronomical observations profit from good image quality. Objects can be better distinguished and structures better resolved as image quality improves. But also the sky background reduces, resulting in better signal-to-noise. Secondly, the La Palma observatory site is known for its very good seeing conditions, similar to other top sites. AO is an excellent way to exploit this advantage. Thirdly, in the visible wavelength range, focal anisoplanatism combined with the small wavefront coherence scale length makes AO on large telescopes technically very demanding. Therefore at large telescopes (e.g. VLT, Keck, Gemini, Subaru) AO developments primarily focus on infrared wavelengths and the existing 4-m class telescopes can complement the large telescopes by focusing on AO at shorter wavelengths. Forthly, LGS systems are being planned for some of the 8-m class telescopes, but fruition of these complex and expensive (multi-conjugate) sodium LGS systems as common-user facilities is likely several years away. This offers a window of opportunity for the 4-m class telescopes to exploit laser systems.

At the WHT there already is an operational AO system (NAOMI), permanently mounted in one of the Nasmyth foci, which has proven capable of delivering excellent image correction at both near-IR and visible wavelengths. The AO-corrected focus is being equipped with a range of instruments: a near-IR imager, INGRID, a coronographic unit, OSCA, and an optical integral field spectrograph, OASIS, which was previously deployed on the Canada-France-Hawaii Telescope. OASIS will facilitate efficient, high spatial resolution 2D spectroscopic observations in the visible wavelength range, where the typical spatial resolution will be 0.2 to 0.3 arcsec. It is this instrument that is expected to deliver the most important scientific advances. For OASIS, the LGS facility will significantly increase the in-lenslet energy, greatly improve the discrimination in crowded science fields when compared to observations without adaptive optics, and dramatically improve the sky coverage and thus the range of science targets that can be studied. The addition of a LGS system to AO instrumentation truly opens AO and in particular OASIS exploitation to a wide range of galactic and extragalactic science.

The fundamental questions being researched in contemporary astrophysics are, in broad terms, related to the formation and evolution of planets, stars, galaxies, and the universe at large. The improved spatial resolution and sensitivity of LGS-AO will have a significant impact on our progress in understanding all of these areas. Whereas in both the science examples highlighted below AO is needed to achieve the necessary spatial resolution and sensitivity, LGS AO is also needed in general to progress from studying only one or two key examples to exploration of samples of objects, and thus truly improve our understanding of the underlying physics.

**LATE STAGES OF STELLAR EVOLUTION AND SUB-STELLAR OBJECTS**

Mostly on the basis of HST narrow-band imaging of young planetary nebulae (PNe) the onset of asymmetry in the outflows from evolved stars is now thought to occur at a much earlier stage than previously expected, namely close to the end of the asymptotic giant branch (AGB). Puzzling systems of knots, filaments and jets have also been discovered in PNe, which contain important information about the dynamics and chemistry in the AGB and post-AGB phases. This makes it especially important to obtain resolved imaging of post-AGB circumstellar envelopes to understand how asymmetry arises and leads to the observed morphological diversity of PNe. Post-AGB envelopes are difficult to resolve from the ground due to their small (<1") angular extent, making AO imaging necessary. OASIS with a LGS system would also allow one to obtain spatially detailed spectral information about the physical conditions, chemical abundances and, for the high velocity outflows, also the dynamical properties of these structures, which will in turn lead to a much better comprehension of basic phenomena such as the mass return of gas to the interstellar medium and its chemistry. In addition, it will offer the ability to measure velocity fields on <1" scales in evolved AGB and post-AGB objects, which will reveal the origin of asymmetry in the velocity field.

The search for companions around solar type stars using precision radial velocities has resulted in a list of more than 70 extra-solar planetary systems orbiting solar type stars. However, a similarly high frequency of brown dwarf (BD) companions around solar-type stars has so far not been detected. The formation scenarios of BDs, which as a population may carry substantial mass in a galaxy, are still unclear but focus on two possibilities: BDs form in orbits around stars and are then ejected, or they form isolated. Recent discoveries of binary BDs[2,3] are in this respect of high scientific interest, but both the observed and theoretically expected frequencies of binary BD systems and of BDs around solar-type and massive stars remain unclear. Equally unclear is whether circum-BD disks exist - finding them would be direct support for a similar formation scenario to that of stars. A LGS AO survey of samples of nearby stars and known BDs is the only way to uncover binary BDs and circum-BD disks.

**CENTRES OF NEARBY GALAXIES**

A key area of study for the LGS-AO system in conjunction with the integral field spectrograph, OASIS, will be the central regions of elliptical and disk galaxies. In particular, attention will be given to aspects relating to the formation of bulges and disks, and to the dynamical interplay between the outer and inner regions of galaxies, which often leads to central black holes and/or nuclear activity.

The intrinsic structure and dynamics of the cores of early-type galaxies can give some of the most direct clues to the processes of galaxy formation and evolution. This is thought to occur either through series of mergers which gradually lead to present-day galaxies, or through monolithic collapse, where elliptical galaxies are formed at early epochs. The central regions hold relics of the formation history of their host galaxies. For example, a merger history can be deduced through the detection of the remnant of a merger with a compact satellite, or through a thick disk in the center, which is a signature of an in-falling system torn apart by tidal forces. A combined approach of observing the stellar populations and the kinematics is necessary, using integral field spectroscopy. Only this 2D spectroscopy allows one to disentangle *spatially* the gaseous and stellar kinematics and star formation parameters like age and metallicity.

In galaxy surveys like the one using the non-AO integral-field spectrograph SAURON, the inner regions of early-type galaxies have been observed[4,5], but at rather low spatial resolution, which precludes observation of the galaxy core in sufficient detail. There, gas accumulates, metal enrichment occurs, dynamical decoupling exists, massive black holes lurk, AGN activity rages, and dynamical signatures of the past history of the galaxy are waiting to be discovered. *HST* imaging has shown that the regions to be studied in this respect really only span the inner few arc seconds. OASIS fed by a LGS AO system will provide unprecedented capabilities by offering the required high spatial resolution and 2D integral field spectroscopy in the visible, giving the required simultaneous measurement of stellar populations, morphology and kinematics. The increase in spatial resolution from typically ~1″ to 0.2″ offered by LGS AO is fundamental. For instance, the radius of dynamical influence of a $10^7$ $M_O$ massive BH is less than 50 parsec, or just under 1″ at 10Mpc distance, thus not resolvable without AO. The great majority of the targets in such studies have no suitable guide sources for conventional AO, hence a LGS is a prerequisite.

# PERFORMANCE EXPECTATIONS

We summarize the key performance characteristics for the Rayleigh LGS system. The central advantage of the proposed LGS system is that high spatial resolution will be attainable over nearly all of the sky. The La Palma observatory site is known for its generally very good seeing conditions (year-round median seeing is 0.69″, and even better during the summer observing season). Site testing has shown that atmospheric turbulence is regularly dominated by low-altitude turbulence[6]. This is an advantage for the LGS system described here, as even with the guide star at relatively low altitude, good illumination of the atmospheric turbulence and hence good wavefront sensing will be achieved.

Wavefront correction performance depends on several parameters. In the table below we summarize results from models, for realistic seeing conditions and associated atmospheric turbulence profiles, based on a laser focused at an altitude of 15km and with a range gate depth of ~100m. Performance predictions are calculated using end-to-end Monte-Carlo simulations, which include representations of the atmosphere, telescope, LGS and AO sub-systems (operating in closed loop) and the science instrument. The model is similar in approach to that described by Ellerbroek[7]. The atmospheric turbulence profile is represented in 3 dimensions as a sequence of "phase screens" which introduce random wavefront phase aberrations with the required spatial power spectrum for atmospheric turbulence (the Tatarski spectrum modified for finite outer scale is assumed). The model calculates the instantaneous phase profile at the telescope entrance aperture for light propagating from sources at arbitrary field angle and range. Time evolution of the phase profile is modeled by horizontal translation of the phase screens (at pre-defined wind velocities) between model iterations.

There is good observational evidence for the localization of atmospheric (optical) turbulence into discrete layers[6]. Here we use a representative 3-layer profile. The layer strengths, heights and wind speeds were chosen to match typical observed profiles for the La Palma site. The atmospheric turbulence ($C_n^2$) profile is modeled as three layers with 40% of the turbulence at the telescope aperture, 40% at a height of 2.5 km and 20% at 7.5 km. Turbulence data are available from a programme of wavefront sensor observations at the WHT obtained over 70 nights[8]. These provide turbulence strength, velocity and (crude) altitude information. Furthermore balloon turbulence profiling[6] and SCIDAR scintillation profiles[9] provided detailed information on turbulence strength versus height and velocity.

| Case | Seeing and $r_0$ | FWHM | | | | D50 | | | |
|---|---|---|---|---|---|---|---|---|---|
| | | R | I | J | H | R | I | J | H |
| **A:** Non-AO natural seeing 0.74″ | 0.74″ / 14cm | 0.71″ | 0.67″ | 0.61″ | 0.58″ | 0.76″ | 0.69″ | 0.61″ | 0.59″ |
| **B:** LGS with *R*=14 NGS on-axis | 0.74″ / 14cm | 0.25″ | 0.13″ | 0.11″ | 0.09″ | 0.56″ | 0.47″ | 0.37″ | 0.31″ |
| **C1:** LGS with *R*=18 NGS on-axis | 0.54″ / 19cm | 0.17″ | 0.14″ | 0.13″ | 0.12″ | 0.39″ | 0.34″ | 0.26″ | 0.23″ |
| **C2:** LGS with *R*=18 NGS on-axis | 0.74″ / 14cm | 0.28″ | 0.19″ | 0.16″ | 0.14″ | 0.56″ | 0.49″ | 0.39″ | 0.33″ |
| **C3:** LGS with *R*=18 NGS on-axis | 0.94″ / 11cm | 0.44″ | 0.33″ | 0.18″ | 0.18″ | 0.73″ | 0.67″ | 0.54″ | 0.47″ |
| **D1:** LGS with *R*=18 NGS 60″ off-axis | 0.54″ / 19cm | 0.21″ | 0.21″ | 0.16″ | 0.15″ | 0.41″ | 0.36″ | 0.28″ | 0.25″ |
| **D2:** LGS with *R*=18 NGS 60″ off-axis | 0.74″ / 14cm | 0.32″ | 0.30″ | 0.20″ | 0.17″ | 0.56″ | 0.54″ | 0.41″ | 0.35″ |
| **D3:** LGS with *R*=18 NGS 60″ off –axis | 0.94″ / 11cm | 0.50″ | 0.40″ | 0.31″ | 0.21″ | 0.75″ | 0.71″ | 0.59″ | 0.48″ |
| **E:** LGS with *R*=19 NGS 90″ off-axis | 0.54″ / 19cm | 0.27″ | 0.23″ | 0.21″ | 0.21″ | 0.42″ | 0.39″ | 0.32″ | 0.28″ |

The 3D atmospheric model allows realistic assessment of the effects of focal anisoplanatism (cone effect) for the laser guide star at finite range, angular anisoplanatism for the natural tilt reference star and for science targets at arbitrary field angles, and also for temporal anisoplanatism - ie. finite servo bandwidth and latency effects in the AO system. Finite outer scale effects are also included. The AO system model includes accurate representations of the segmented deformable mirror, the fast steering mirror, Shack-Hartmann wavefront sensor, tilt sensor, and the closed loop control system. The effects of DM fitting errors (finite spatial order of correction), photon noise, detector readout noise, and seeing (speckle "noise") in the wavefront sensors are all included. The effects of focus error and LGS WFS spot elongation are not included in the simulations and error budgets. However, the magnitudes of these errors have been estimated, indicating that the effects of these errors will be small for the LGS range of 15 km.

LGS operation still requires the presence of a natural guide star in order to correct for overall wavefront tilt. Although this natural source can be much fainter and at a larger distance from the science object than with normal, non-LGS AO operation, it still poses some limitations on the quality of image correction. Therefore we calculated also situations with off-axis natural guide stars.

Values of FWHM and diameter of 50% encircled energy (D50) are presented for different seeing conditions, guide star magnitudes, distances of the natural guide star to the science object, and wavelengths. The simulations were performed at wavelengths corresponding to *R*, *I*, *J* and *H* bands. Cases B and C1,2,3 present realistic cases for bright and faint self-referencing objects, indicating important improvements in image quality for objects as faint as *R*=18. Hence for *any* point source down to *R*=18 good AO correction will be obtained. For fainter science targets or diffuse sources an off-axis natural guide star is required. Cases D1,2,3 in the table refer to a faint guide star that is 60″ away from the science object, implying full sky coverage in the galactic plane, 90% at 30° galactic latitude, and 50% coverage at the galactic pole. For case E, where the natural guide star is very faint and even further off axis, ~100% sky coverage is obtained (>90% at the galactic pole). For comparison, AO at visible wavelengths without a laser guide star provides a sky coverage of only <1%. The field diameter over which the corrections will be maintained is of order 1′ (in the *I* band under median seeing conditions) and hence is well matched to the field available to the instruments.

Comparing case A for natural seeing with cases B, C2 and D2 indicates the significant improvements that can be expected under typical atmospheric conditions. Even in the extreme case E for virtually full sky coverage the image quality improvements are still very attractive. Although high Strehl ratios will not be achieved at visible wavelengths, the improved image sharpness still implies an important advantage in in-slit energy for a spectrograph such as OASIS over the natural, uncorrected seeing.

An example point spread function at 850nm for case C1 in the table above is compared with the natural, uncorrected seeing case in Figure 1. Corresponding Strehl ratios are 0.09 (uncorrected) and 0.22 (corrected).

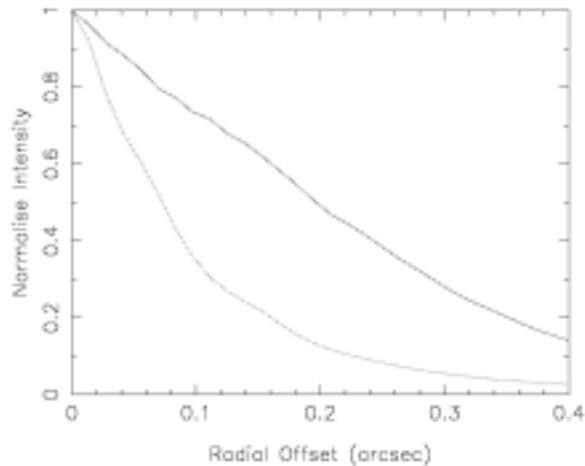

Figure 1: Azimuthally averaged point spread function for the AO-corrected and uncorrected case at 850nm.

Based on seeing statistics for La Palma we anticipate that appropriate conditions for LGS operation exist about 3/4 of the available observing time[10]. On occasions, scattering by dust or thin cirrus in the atmosphere may hamper efficient laser operation, but this is expected to interfere with observations only a small fraction of the time.

# TECHNICAL DESCRIPTION

**OVERVIEW**

There are essentially two different methods for generating an artificial laser beacon: (i) a laser beam tuned exactly at the wavelength of a transition that excites a layer of sodium atoms at an altitude of about 90 km, or (ii) using Rayleigh back scattering of a pulsed laser beam to produce a point source at a lower altitude. The first option in principle would deliver the best possible AO correction, but currently requires expensive laser technology with high operating cost. The latter option is achievable with existing, relatively inexpensive, commercially available laser technology and would still deliver the objective of high sky coverage and good AO correction. For reasons of cost effectiveness and speed of development we have chosen to pursue a Rayleigh LGS for the WHT.

The Rayleigh laser system is designed to work in conjunction with existing AO equipment and ancillary instrumentation and infrastructure. A 25W pulsed frequency-doubled Nd:YLF (or YAG) laser, emitting at 523 (or 532) nm, will be projected at an altitude of at least 15km from a launch telescope mounted behind the secondary mirror. The Rayleigh back-scattered light will be detected by a wavefront sensor system (LGS-WFS) to measure the wavefront shape from the LGS, and provide corrections to the existing deformable mirror of the AO system. A Pockels cell range-gate system that is synchronized with the laser pulses will set height and duration of the laser return beam. The existing natural guide star wavefront sensor system (NGS-WFS) will be dedicated to tip-tilt correction measurements using a nearby star, or, if suitable using the science target itself. The common-user Rayleigh LGS facility will comprise the following sub-systems:

- Beam Launch Telescope (BLT) – comprising: the laser head, controller and cooler, beam expansion, collimation and focusing optics, beam steering mechanism (to compensate for telescope alignment and flexure), beam diagnostics (beam profile and power monitoring), and a bore sight camera for alignment.

- Laser Guide Star Wavefront Sensor (LGS-WFS) – comprising: dichroic pick-off of the return laser light, Pockels range-gate shutter, optics, CCD and controller, and interface to the NAOMI control system.

- An upgrade of the existing NAOMI Natural Guide Star WFS (NGS-WFS) with a very low-noise detector to optimize closed loop tip-tilt operation on fainter stars.

- Enhancement of the NAOMI software suite and real-time control system to accommodate LGS operation, and laser safety equipment.

**LASER PROJECTION**

The Rayleigh LGS will be used at zenith angles up to 60°, allowing several hours of on-target observations and making all of the northern and a large part of the southern hemisphere accessible. The laser beacon height is set at 15 km. Performance predictions indicate limited advantages in operating at longer ranges and the final range is constrained by the available laser power and the desired atmospheric photon return. Although variable-range operation could be an advantage in some circumstances, fixed-range operation reduces the complexity and cost of the beam launch telescope, the LGS wavefront sensor and the associated software development.

For a 25W laser operating at 7kHz, the energy per pulse is 3.6mJ. Estimated photon returns, taking into account atmospheric density, scattering cross section, and overall system transmission and detection efficiency, and assuming a 100m range gate depth predict a detection of 8 photo-electrons per laser pulse per Shack-Hartmann sub-aperture. A 3ms WFS exposure will encounter 21 pulse returns from sky, yielding some 170 photo-electrons per subaperture per integration. Modeling predictions indicate that an AO system based on 6x6 pixels per WFS subaperture with $5e^-$ readout noise can operate successfully at levels of 100 or more photons per subaperture per integration.

The diameter of the BLT output beam will be approximately 40 cm in order to produce a sufficiently small spot size at 15 km. The BLT must correct for misalignment, telescope flexure, thermal effects, atmospheric effects and vibration.

The preferred conceptual design is based on a refractive telescope for optimum alignment stability. A beam steering mirror will be provided in the BLT to stabilize the LGS position, driven by the tip-tilt error signals from the LGS-WFS. Diagnostics to monitor the laser beam quality and power will be included.

**WAVEFRONT SENSING**

The back-scattered light from the laser beam will be transmitted through the WHT and into NAOMI's optical train. A schematic layout of the existing NAOMI AO system together with the additional optics for the LGS-WFS is shown in Figure 2. A dichroic mirror will pick off the laser light after it has passed the deformable mirror, and direct it towards the Pockels cell shutter, the lenslet array, and the LGS-WFS. The dichroic mirror will be the only additional optical element in the light beam towards the science focus, and hence have little negative impact on the science capability.

Re-imaging optics will provide the correct beam size. Light at all other wavelengths will pass the dichroic pickoff and pass on towards the NGS-WFS and the science cameras.

The mean range and depth (vertical extent) of the LGS will be set by an electro-optical range gate unit acting on the back-scattered radiation entering the LGS-WFS. This shutter system will be based on two large aperture Pockels cells (operating between crossed polarizers, one for each polarization state) in order to achieve the desired speed of operation and 'closed' extinction ratio.

The WFS design will be based on an 8 by 8 lenslet Shack-Hartmann sensor with a Marconi 80 x 80 pixel CCD39 that will sense the high-order wavefront distortions.

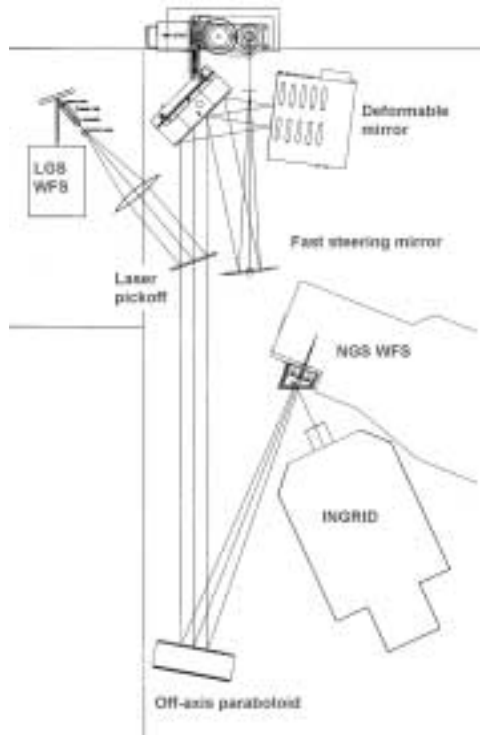

Figure 2: Conceptual drawing of the LGS-WFS optical path in conjunction with the NAOMI optical train. This system is mounted on an optical bench in the Nasmyth focus of the WHT. The telescope focus is located at the top of the drawing. The space envelope of the existing NGS-WFS is indicated. INGRID is the IR science camera.

The monostatic launch approach, with the laser being launched from behind the secondary mirror as opposed to alongside the telescope, has the advantage of (i) minimizing laser spot elongation, and (ii) avoiding problems of the laser spot rotation relative to the wavefront sensor camera. With a 100m range gate depth the subapertures at the edge of the WFS view a spot elongation of only ~0.2″, which is much smaller than the intrinsic spot size of ~1″ due to diffraction of the launch telescope optics, atmospheric turbulence, and laser beam quality.

Wavefront tilt information must be obtained from a natural guide star. The existing NAOMI NGS-WFS with its star pickup arrangement will fulfill this function. This sensor will be required to operate with natural guide stars down to at least magnitude $R$=18 or fainter, to provide high sky coverage. A E2V L3CCD zero read noise CCD will be installed for this purpose. The upgraded NGS-WFS will operate on stars within 90″ radius of the science objects.

## FOCUS ERRORS

The finite range and depth of the LGS contributes additional curvature to the wavefront seen by the LGS WFS system. This curvature is a source of focus error that must be taken into account, i.e., one must be able to distinguish between focus changes caused by atmospheric turbulence and those attributable to the LGS.

The LGS is not a point source but an illuminated column within the atmosphere. Thus the LGS can be regarded as a series of sources distributed axially and transversely, with the axial distance (depth) greatly exceeding the transverse dimensions. These sources produce spherical wavefronts, subsequently degraded by atmospheric turbulence, with slightly different curvatures at the WHT entrance aperture. Only one curvature can be removed from the sum of these phases. Ideally the average wavefront curvature attributable to the LGS should be fixed. There are two effects that can cause this curvature to vary. Range gating will be used to set the effective range of the Rayleigh LGS, but timing variations in the range gating will result in range variations that lead to focus errors. Furthermore, if there are time-dependant variations in the backscatter strength at different ranges within the illuminated column then this variation will also lead to a focus error.

A Rayleigh LGS offers advantages over a LGS in the mesospheric sodium layer in that the depth of the backscatter region is much smaller, (e.g. 100 m vs. about 10 km for the sodium layer) and under clear conditions the Rayleigh LGS backscatter will not exhibit the variability of the sodium return. The vertical distribution of the atoms in the sodium layer is known to vary considerably, and occasionally the layer can effectively divide in two[10]. Although the greater range of the sodium-layer LGS helps to offset the effect, the errors can still be significant. As a result, focus sensing with a NGS is usually required for sodium-layer LGS systems.

To illustrate the magnitude of the focus error introduced by an uncertainty in the effective range of the LGS due to range gate timing "jitter" and/or altitude-dependent backscatter variations, we calculate the wavefront focus change over the telescope aperture as r.m.s. waves at the laser wavelength. These values are inversely proportional to wavelength and thus the defocus effects will be smaller for science instruments that use longer wavelengths, as is usually the case. Expectations are that the range gate timing jitter will be below 50 ns peak-to-peak, corresponding to a range uncertainty of about 8 m, taking into account the double pass of the light to and from the LGS. The corresponding time-averaged wavefront focus error should be less than 0.05 waves r.m.s. at the laser wavelength.

The finite depth of the LGS also introduces a defocus effect as only one cross-section of the LGS image can be in perfect focus at the WFS CCD. This effect is far smaller than the degradation due to atmospheric turbulence.

## ERROR BUDGET

The top-level error budget presented here focuses on the in-lenslet energy in an 0.3″ wide hexagonal lenslet of the OASIS integral field spectrograph. The high sky coverage case at a wavelength of 850 nm is taken as a representative example for typical operating conditions. Figure 3 shows this cumulative error budget for the in-lenslet energy for OASIS. The reduction factors (RF) shown are referenced to the diffraction-limited performance. At a wavelength of 850 nm the diffraction-limited case provides an in-lenslet energy of 91%, whilst the uncorrected case would yield only 12%.

The cumulative error budget has the fortuitous result that tilt anisoplanatism has little effect for OASIS. The point spread function is already fairly broad before introducing the tilt anisoplanatism and thus there is little effect on the in-lenslet energy. Likewise, cross-talk between adjacent lenslets is hardly affected. However, the effect of tilt anisoplanatism would be significant for an application requiring a high Strehl ratio at a short wavelength, which could only be achieved in very favorable turbulence profile conditions.

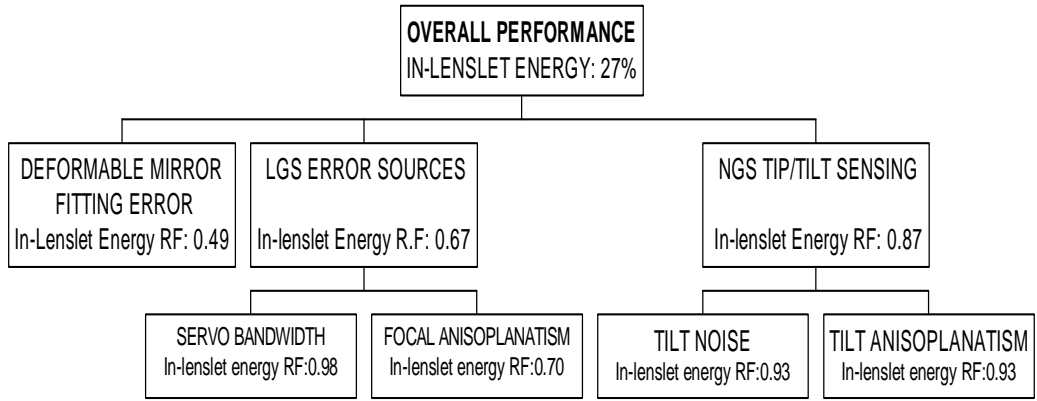

Figure 3: Top-level error budget for OASIS lenslet energy performance.

## OPERATIONAL PROCEDURES, INFRASTRUCTURE, AND SAFETY SYSTEMS

Normal night-time operation of the laser will be fully by remote control from the telescope control room. The LGS facility will be designed for high reliability and ease of use. AO observations involve a complex interplay of atmospheric conditions, availability of natural guide stars, astronomical programme constraints and instrumentation, all affecting the overall observing efficiency and success of scientific experiments. Typical overhead for field setup, acquisition and lock-on with NAOMI is less than 5 minutes and this should reduce even further with LGS use because a bright point source for high-order correction will readily be available.

AO observations will principally be carried out in queue-scheduled mode, biased toward the best seeing summer months, so that the best seeing periods will be exploited for AO. To work effectively, independent seeing measures must be available, and to that effect a robotic differential image motion monitor is under construction.

Safety aspects focus on personnel safety systems to prevent accidental exposure to bright laser light. The unexpanded laser beam will greatly exceed the safe levels for skin exposure. Even the expanded beam is a significant eye hazard. An aircraft detection and interlock system is also planned to prevent illumination of passing aircraft, even though the skies above the observatory are a no-fly zone. Furthermore, interference with other telescopes will be prevented by a cross-telescope beam collision interlock system.

## PROSPECTS

The system description as presented here in outline is based on a study performed as part of proposal for funding. If approved, we expect to have an operational system within 2½ years from project initiation. Apart from the intrinsic science potential our LGS system would deliver, it would also allow engineers and astronomers to gain experience with the practical use of laser guide stars. Furthermore, the system could serve as a test bed for future advanced laser systems such as required for multi-conjugate AO experiments.

## REFERENCES

1. Sandler D. G., 1999, in "Adaptive Optics in Astronomy", Cambridge University Press, ed. F. Roddier, p. 331
2. Martin E. L., Brandner W., Basri G., 1999, Science **283**, 1718
3. Close L. M., Potter D., Brandner W., Lloyd-Hart M., Liebert J., Burrows A., Siegler N., 2002, ApJ **566**, 1095
4. Bacon R., Copin Y., Monnet G., Miller B. W., Allington-Smith J. R., Bureau M., Carollo M. C., Davies R. L., Emsellem E., Kuntschner H., Peletier R. F., Verolme E. K., de Zeeuw T. P., 2001, MNRAS **326**, 23
5. de Zeeuw P. T., Bureau M., Emsellem E., Bacon R., Carollo M. C., Copin Y., Davies R. L., Kuntschner H., Miller B. W., Monnet G., Peletier R. F., Verolme E. K.,, 2002, MNRAS **329**, 513
6. Vernin J., Muñoz-Tuñón C., 1994, A&A **284**, 311
7. Ellerbroek B. L., 2001, in proc. ESO conf. "Beyond Conventional Adaptive Optics", eds. R. Ragazzoni, N. Hubin and S. Esposito.
8. Wilson R. W., 1998, New Astronomy Reviews **42**, 465
9. Klueckers V. A., Wooder N. J., Nicholls T. W., Adcock M. J., Munro I., Dainty J. C., 1998, A&AS **130**, 141
10. Wilson R. W., O'Mahony N., Packham C., Azzaro M., 1999, MNRAS **309**, 379
11. Michaille L., Clifford J. B., Dainty J. C., Gregory T., Quartel J. C., Reavell F. C., Wilson R. W., Wooder N. J., 2001, MNRAS **328**, 993